\documentclass[useAMS,usenatbib]{mn2e}
\usepackage{graphicx}

\def\lesssim{\,\lower2truept\hbox{${<\atop\hbox{\raise4truept\hbox{$\sim$}}}$}\,}
\def\gtrsim{\,\lower2truept\hbox{${>\atop\hbox{\raise4truept\hbox{$\sim$}}}$}\,}

\title[The growth of the nuclear BH in SMGs]
{The growth of the nuclear black holes in submillimeter galaxies}
\author[Granato et al.]{G.L. Granato$^{1,3}$, L. Silva$^{2}$,
A.\ Lapi$^{3}$, F.\ Shankar$^{3}$,  G. De Zotti$^{1,3}$ and L.
Danese$^{3,1}$\\
$^{1}$INAF - Osservatorio Astronomico di Padova\\
$^{2}$INAF - Osservatorio Astronomico di Trieste\\
$^{3}$SISSA}
\begin{document}



\maketitle

\label{firstpage}

\begin{abstract}
We show that the ABC scenario we proposed for the co-evolution of
spheroids and QSOs predicts accretion rates and masses of
supermassive black holes in sub-mm galaxies in keeping with recent
X-ray determinations. These masses are well below the local
values, and those predicted by alternative models. The observed
column densities may be mostly due to ISM in the galaxy. The
contribution of the associated nuclear activity to the X-ray
background is likely negligible, while they may contribute a
sizeable fraction $\sim 10 \%$ to hard-X cumulative counts at the
faintest observed fluxes.\end{abstract}

\begin{keywords}
galaxies: formation – galaxies: active – cosmology: theory -
X-rays: galaxies
\end{keywords}
\section{Introduction}

In recent years, many efforts have been devoted to understand the
population of high-z galaxies detected by sub-millimeter surveys
(SMGs), which could dominate the $z>2$ cosmic star formation (SF),
and may pinpoint the major epoch of dust-enshrouded spheroid
formation, as suggested by the their SF level, high mass fraction,
large dynamical mass and strong clustering (Smail et al.\ 1997,
Hughes et al.\ 1998; Barger et al.\ 1998; Blain et al.\ 2004;
Greve et al.\ 2005).

From the theoretical side, the $\Lambda CDM$ cosmology is a
well-established framework to understand the hierarchical assembly
of dark matter (DM) halos, but the complex evolution of the
baryonic matter remains an open issue, because whatever simulation
must include huge simplifications for its physics. Unfortunately,
these simplifications pertain processes which are major drivers of
galaxy evolution, such as SF, feedback and nuclear accretion.

The class of computations known as semi-analytical models (SAM)
have been extensively compared with a large range of observations
at various redshifts. Besides many successes, some difficulties
persist in standard SAM, broadly connected with massive galaxies
(e.g. the color-magnitude relation, the $[\alpha/Fe]-M$ relation,
the statistics of sub-mm and deep K-band selected samples: see
Thomas, Maraston, \& Bender 2002;  Pozzetti et al. 2003;
Sommerville 2004; Baugh et al 2005, Nagashima et al 2005).
However, the general agreement of a broad variety of data with the
hierarchical scenario for DM and the fact that the observed number
of luminous high-redshift galaxies, while substantially higher
than predicted by standard SAMs, is nevertheless consistent with
the number of sufficiently massive DM haloes, indicate that we may
not need alternative scenarios, but just some new ingredients or
assumptions for visible matter.


In Granato et al. (2001) we  suggested that a crucial ingredient
to keep into account is the mutual feedback between spheroidal
galaxies and the AGN at their centers, largely ignored by
simulations at that time, and often even now. This despite the
fact that since long observation have suggested a connection
between SF and AGN activity (e.g.\ Sanders et al.\ 1988), the
possible importance of the feedback from the central BH has been
discussed (among the first Ciotti \& Ostriker 1997; Silk \& Rees
1998; Fabian 1999), and its growth  has been considered in models
for galaxy formation (e.g. Kauffman \& Haenelt 2000; Archibald et
al. 2002). In Granato et al. (2004, GDS04 hereafter) we presented
a physical model for the early co-evolution of the two components,
in the framework of the hierarchical $\Lambda CDM$ cosmology and
based on the semi-analytic technique. In our model, the SMGs are
interpreted as spheroids observed during their major episode of
star-formation (SF). The development and duration of this episode
is affected not only by supernova (SN) feedback, but also by the
growth by accretion of a central super massive black hole (SMBH),
{\it favored by the SF itself}, and by the ensuing feedback by QSO
activity, and completes earlier in more massive
objects\footnote{Thus we named our scenario {\it Anti-hierarchical
Baryon Collapse - ABC}; from the observational point of view the
same phenomenon is now commonly referred to as {\it down-sizing}}.
Thus the high redshift QSO activity marks and concur to the end of
the major episode of SF in spheroids.


The scenario by Granato et al. (2001) and GDS04 is based on a
circular relationship between SF and AGN activities, which
establishes a well defined sequence connecting various populations
of massive galaxies: (i) virialization of DM halos; (ii) vigorous
and rapidly dust-enshrouded star formation activity, during which
a central SMBH grows; (iii) QSO phase halting subsequent star
formation and (iv) essentially passive evolution of stellar
populations, passing through an Extremely Red Object (ERO) phase.
As detailed by GDS04 and Silva et al.\ (2005), this scenario fits
nicely two important populations at high redshift, which are
instead problematic for most semi-analytic models (e.g.\
Sommerville et al. 2004): vigorously star-forming, dust-enshrouded
starbursts (in practice SMG; stage (ii)) and ERO (stage iv). Also,
the local luminosity function of spheroids and the mass function
of SMBHs are well reproduced. The general consistence of this
sequence with high redshift QSO population has been investigated
by Granato et al.\ (2004), while a detailed analysis will be
presented by Lapi et al.\ (in preparation).

In this paper we focus on the model behavior during the SMBH
growth in stage (ii), as traced by X-ray observations of sub-mm
selected sources. Alexander et al. (2003, 2005a, 2005b) have
investigated the X-ray properties of bright SMGs (S$_{850 \mu m}
\gtrsim 4$ mJy), by combining ultra-deep X-ray observations (the 2
Ms CDF-N) and deep optical spectroscopic data of SMGs. They have
found evidence for the presence of mild AGN activity in a large
fraction of bright SMGs ($\sim 50$\%), which in our scheme is the
signature of this growth, that afterwards will quench the SF, and
cause an almost passive evolution of stellar populations.

Recently sub-grid treatments of SMBH growth and its feedback has
been implemented in numerical simulations of DM halos and
large-scale SPH gas dynamics (e.g.\ Springel, Di Matteo, \&
Hernquist, 2005), or in the semi-analytic post-processing of the
Millenium DM simulation (Croton et al.\ 2005, Bower et al.\ 2005).
However, in the latter cases, the feedback role of AGN is limited
to the 'radio mode', which suppress cooling flows in clusters at
z\lesssim 1.


\section{The GDS04 model} \label{sec:model}

This paper is based on the SAM presented by GDS04, which follows
the evolution of baryons within proto-galactic spheroids through
simple but physically grounded recipes.  We provide a qualitative
summary of the model, deferring the reader to that paper for
details.

The treatment of the statistic of DM halos essentially follows the
standard framework of hierarchical clustering.

During the formation of the host DM halo, the baryons are
shock-heated to the virial temperature; this hot gas cools fast
especially in the dense central regions or clumps, and triggers a
huge burst of star formation. One of the major differences with
respect to most semi-analytic treatments of the evolution of
visible matter, is that in GDS04 cool and collapsing gas forms
stars without setting in a quiescent disc.

\begin{figure}
\centering
\includegraphics[angle=0,width=8cm,height=10.0cm]{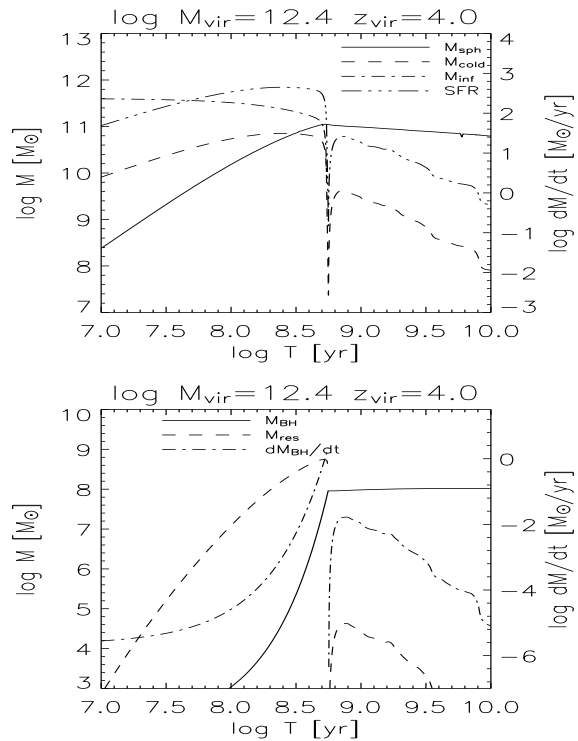}
\caption{Top panel: evolution of the stellar mass, infalling mass,
cold gas mass, and star formation rate within a typical halo of
total mass $2.5\times 10^{12} M_\odot$, virialized at redshift
$4$. Bottom panel: evolution of the BH mass, reservoir mass, and
BH accretion rates in the same halo. The discontinuity at $\log T~
\sim 8.75$ marks the epoch at which the AGN feedback sterilize the
system.} \label{fig_sfr_bh}
\end{figure}

The SF activity promotes the gathering of a reservoir of gas with
angular momentum low enough to allow accretion onto a nuclear
SMBH. A plausible process is radiation drag, which 'naturally'
yields a ratio between SMBH and spheroid mass in keeping with
observations. The reservoir gas eventually accretes onto the SMBH,
powering nuclear activity. The maximum accretion rate is of the
order of the Eddington limit, so that the SMBH mass and the BH
activity increases exponentially with time. The energy feedback to
the gas by SN explosions and BH activity affects the ongoing SF
and BH growth. The two feedbacks have very different dependence on
halo mass and on time since virialization. The SN feedback is an
almost immediate consequence of SF and its time evolution reflects
that of SF. It is very effective in low mass halos, severely
limiting the growth of stellar and SMBH component, but it is of
minor importance in most massive galactic halos; the AGN feedback
grows exponentially, it is negligible in the fist $\sim 0.5$ Gyr
in all halos, but suddenly becomes very important in DM halos
massive enough to be little affected by SNae feedback. Thus in low
mass halos, SNae keep the SF at low level, also limiting the
growth of the SMBH and its capability to influence the system. In
moderate to high mass halos (see Fig \ref{fig_sfr_bh}), the SNae
becomes increasingly less effective, the SMBH can grow efficiently
and after a time delay $\sim 0.5$ Gyr, required by the exponential
growth, quenches any further substantial SF, and ultimately its
own activity. Since after this point the SMBH is already present
at the galaxy center, any subsequent supplying of gas (e.g. due to
merging or accretion of the halo) produces an immediate AGN
feedback, and thus is unable to alter substantially the star and
SMBH mass: the stellar populations evolve largely in a passive
way.

Before the peak of the accretion, the SMBH is likely to be
obscured by the surrounding galactic ISM, therefore it could
possibly be detected only in the hard-X rays while the
proto-galaxy appears as a SMG: {\it the present paper is precisely
devoted to study this pre-qso phase}. Later on, in the proximity
of the peak, i.e., when the central SMBH is powerful enough to
remove most of the gas and dust from the surroundings, the system
will shine as an optical quasar.

\section{The accretion of SMBHs in SMGs}

The GDS04 model predicts the bolometric {\it intrinsic} time
development of AGN activity in forming spheroids. However detailed
computations on how and when this activity may show up are made
extremely uncertain in most electromagnetic bands by environmental
effects. For instance, optical-UV bands are heavily affected by
obscuration due both to the general galactic ISM and to that
around the very central region. In the IR region, where
obscuration is much less a problem, it is however difficult to
disentangle dust emission powered by the AGN from that related to
SF activity. This is particularly true in our scenario, since the
BH growth occurs in a extremely dust-enshrouded ambient, without
obvious analogue in the local universe. The situation is more
favorable with X-ray photons, especially hard X ones, which are
less affected by interactions with the ISM, and are also less
likely to be confused with those produced by processes connected
with SF, such as X-ray binaries.

\begin{figure}
\centering
\includegraphics[angle=0,width=8cm]{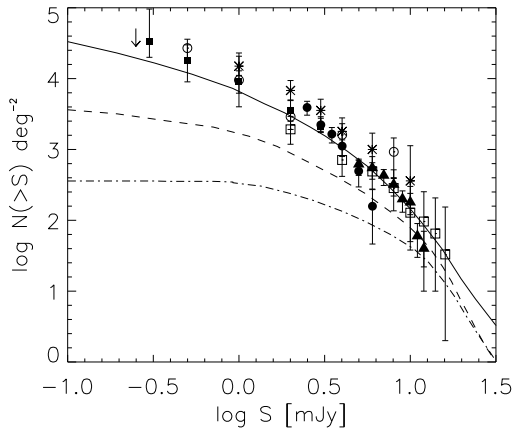}
\caption{Number counts predicted by the GDS04 model for SCUBA
sources with accretion rates greater than two interesting
thresholds. Solid: all sources; dashed line: $\dot M > 0.013
M_{\odot} \mbox{yr}^{-1}$ ($L_{\rm bol} \gtrsim 10^{43} \mbox{erg
s}^{-1}$); dot-dashed line: $\dot M
> 0.13 M_{\odot} \mbox{yr}^{-1}$ ($L_{\rm bol} \gtrsim 10^{44}
\mbox{erg s}^{-1}$). Data from Blain et al. (1999; open circles
and upper limit), Eales et al. (2000; filled circles), Chapman et
al.\ (2002; asterisks), Cowie et al.\ (2002; filled squares),
Scott et al. (2002; filled triangles), Borys et al.\ (2003; open
squares).}
\label{fig_counts}%
\end{figure}


Fig.\ \ref{fig_counts} shows number counts (see e.g.\ De Zotti et
al.\ 1996 for definitions) for SCUBA sources hosting accretion
rates onto the central SMBH greater than two interesting
thresholds. We predict that about 40\% and  15\% of SCUBA sources
brighter than $\simeq$ 5 mJy have $\dot{M}_{BH}$ greater than
$\sim$ 0.013 and 0.13 $M_{\odot}$ yr$^{-1}$ respectively (dashed
and dot-dashed lines in Fig.\ \ref{fig_counts}). With the
accretion efficiency 0.15 assumed in the model, and adopting a
plausible bolometric correction of $L_{\rm bol}/L_{X}[0.5-8 {\rm
keV}] \simeq 17$ (Elvis et al 94, Marconi et al.\ 2004), these
values translates to an accretion {\it intrinsic} luminosity
$L_{X}[0.5-8 {\rm keV}]$ of $\sim 10^{43}$ and   $10^{44}$ erg/s
respectively. These figures compare very well with the findings by
Alexander et al. (2003, see also Alexander et al.\ 2005a,b). They
found that a fraction $30-50\%$ of bright ($\gtrsim 5$ mJy) SCUBA
sources hosts mild AGN activity, with X ray (0.5-8 keV) {\it
intrinsic} luminosity between $10^{43}$ and $10^{44}$ erg
s$^{-1}$. This fraction raises to $\sim 70\%$ in the radio
selected SMG sample studied by Alexander et al.\ 2005a,b. However
they caution that this sample may have an higher incidence of AGN
activity than the whole SMG population, due to the condition of
radio and spectroscopic identification. Anyway, the observed high
fraction of SMGs harboring mild AGN activity indicates that they
are accreting constantly, hence supporting the picture presented
here.


\begin{figure}
\centering
\includegraphics[angle=0,width=8cm,height=5.5cm]{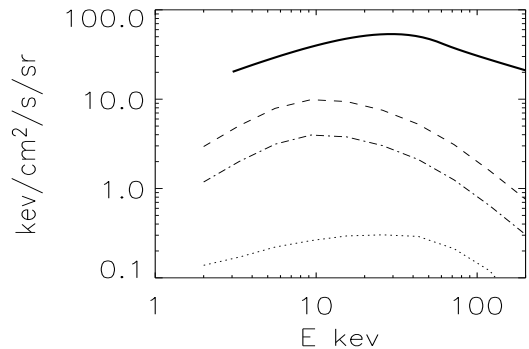}
\caption{Predicted X-ray background from growing SMBH in all
forming spheroids (short dashed line), and from those in SCUBA
sources brighter than 1 mJy (dot-dashed line). We adopted $\log
N_H =23.5$ (see discussion in the text). The dotted line is the
estimated contribution from stellar populations computed with
GRASIL (Silva et al.\ 1998, 2003). The data (thick solid line),
are those adopted by Ueda et al (2003).}
\label{fig_xrb}%
\end{figure}

\begin{figure}
\centering
\includegraphics[angle=0,width=8cm,height=5.5cm]{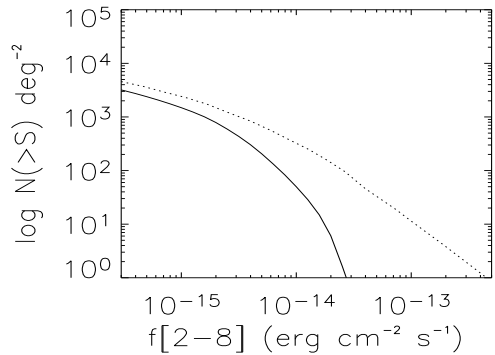}
\caption{Contribution to the X-ray number counts by growing SMBH
in forming spheroids according to the model (solid line). In the
computation we have adopted a reference value of $\log N_H =23.5$
(see discussion in the text). The dotted line outline the observed
counts from various sources (Gilli 2003).}
\label{fig_cx}%
\end{figure}

\section{Estimates of column densities} \label{sec:column}

According to our interpretation, the moderate AGN activity
revealed by X-ray observations in many bright SCUBA sources
corresponds to the build up by accretion of the central SMBH,
induced by star formation, and well before the bright QSO phase
that causes the end of the major epoch of star formation in these
objects.

Alexander et al. (2003, 2005a, 2005b), from the fitting to the
X-ray spectra, derive a column density to the central AGN in the
range N$_H \sim 10^{20}- 10^{24}$ cm$^{-2}$, most sources showing
N$_H \gtrsim 10^{23}$ cm$^{-2}$, and a column-density distribution
roughly similar to that found for nearby AGNs. Hints on the
presence of even Compton-thick absorption in these sources are
given by the $\sim 1$ kev equivalent width Fe K$\alpha$ emission
line seen in the composite X-ray spectrum of the most obscured
objects in their sample.

It is interesting to note that, according to our model
interpretation, the values of column densities to the nucleus
could be dominated by the general ISM of the galaxy alone, rather
than by an obscuring torus such as those invoked by unified models
for broad lined and narrow lined AGNs. Indeed during the bright
SCUBA phase and after a SMBH sufficiently massive to explain the
observed intrinsic X-ray luminosities has developed (i.e. BHs with
mass in the range M$_{BH} \simeq 5 \, 10^5 - 10^7$ M$_\odot$ have
accretion rates $\dot M_{BH} \simeq 0.01 - 0.1$ M$_\odot$/yr, see
Fig. \ref{fig_counts}), the gas mass in galaxies with S$_{850 \mu
m} \geq 5$ mJy (corresponding to M$_{vir} \gtrsim 2 \, 10^{12}$
M$_\odot$) is usually enough to produce column densities to the
nucleus $N_H \gtrsim \mbox{a few} \times 10^{23}$ to a few $\times
10^{24}$. This is the case if the radial density distribution is
sufficiently centrally concentrated, such as a typical King
profile with core radius $\sim 0.3-1$ Kpc (depending on the galaxy
mass). However, the precise value of $N_H$ is strongly dependent
on the assumed profile. A lower limit is obtained if the gas is
more or less uniformly distributed in the galaxy. In this case
$N_H$ would drop by 2-3 orders of magnitudes, becoming almost
negligible with respect to the observed estimates. Note also that
the observed values of $N_H$, when translated to optical dust
absorption adopting a standard conversion factor ($N_H\sim 1.5
\times 10^{21} A_V$), yield large values, consistent with the fact
that these AGN are unseen at optical wavelengths.

In GDS04 the fuelling of the central BH in SMGs takes place in two
steps: (i) a low-angular momentum gas reservoir is formed, that
(ii) then is accreted onto the BH in a timescale set by the
Eddington limit (see Sec. \ref{sec:model}). Thus the growing BHs
may also be hidden by the reservoir. If we assume that its gas is
distributed within a region of $\sim 100-200$ pc, i.e.\ the
typical size of the dusty torii explaining the IR SED of type 1
and 2 AGNs (e.g. Granato \& Danese 1994), then the N$_H$ we obtain
during the SMG phase are between a few $10^{23}$ to $10^{26}$
cm$^{-2}$. These values are again compatible with the
observational estimates. If the matter is distributed around the
BH in a toroidal-type shape, rather than completely surrounding
it, the visibility of the central AGN would depend also on the
distribution of the aperture angles.

\section{The masses of SMBHs in SMGs}

Recently, some hydrodynamic simulations of galaxy major mergers
incorporated, with sub-grid approximations, the growth of the
black hole and its feedback on the evolution of the system (e.g.,
Di Matteo et al. 2005; Springel et al. 2005). As noticed by
Alexander et al.\ (2005) the black holes of the most massive
galaxies in these simulations (which may represent in this merging
scenario the SMGs) are up to an order of magnitude more massive
than those estimated in SMGs, under the assumption of Eddington
limited accretion, and confirmed by the relative narrowness of
broad emission lines detected in some sources. These estimates are
instead in good agreement with our predictions, at least on
average.

This basic difference is due to the fact that in the merging
scenario the most active SF phase, corresponding to the final
merge, is preceded by a long $\sim 1$ Gyr phase of disturbance
which causes a substantial growth of the SMBH. As a result, when
the final merge occurs, the SMBH is already massive enough to
immediately accrete all the matter funnelled in its proximity. By
converse, in our scenario the SFR reaches levels close to the peak
value on a short timescale $\lesssim 0.1$ Gyr, and remains at
these levels until the Eddington limited growth of the SMBH, which
requires $\sim 0.5$ Gyr to build up the final mass, ultimately
sterilize the system (see Fig \ref{fig_sfr_bh}). Then in this case
during most of the "burst" the SMBH is well below its final mass.

Borys et al.\ (2005) have investigated the relationship between
the SMBH and stellar mass in SMGs $M_{BH}/M_{*}$, finding values
1-2 orders of magnitude smaller than those of local spheroids with
similar masses. They notice that this result may be affected by a
few assumptions, and their sample has a modest dynamic range
($\log M_*=11.4 \pm 0.4$). In our scenario we would expect, during
the SMGs phase, an average $M_{BH}/M_{*}$ of the same order of
magnitude as that found by Borys et al, but with a larger
dispersion. The observed lack of detected objects with smaller
values may be easily accounted for by selections effects, due to
the low accretion rate (and possibly high obscuration) of AGN
activity in the very early phases. As for the lack of observed
SMGs with $M_{BH}/M_{*}$ closer to the local value, the more
obvious possibility is that an important ejection of the dusty ISM
begins somewhat earlier than what our schematic model predicts,
causing a decrease of the sub-millimeter flux below the current
sensitivity in the last 3-4 e-folding times of SMBH growth before
the maximum. This would not affect too much the good match with
observed SMGs counts, while explaining the lack of sub-millimeter
sources with more evolved SMBH. Note also that the reservoir mass
increases steadily up to very close to the maximum of the
accretion rate (see Fig \ref{fig_sfr_bh}). If this reservoir were
responsible for most of the obscuration (see section
\ref{sec:column}) the last e-folding times of the SMBH growth
before the optical QSO shining could be undetectable even in the
X-ray band, because the increasing optical depth would overwhelm
the increasing intrinsic power. Deeper sub-mm and X-ray data will
clarify the issue.


\section{X-ray background and counts}

A detailed prediction of the contribution of forming spheroids and
SMBH to X-ray counts and background (XRB), in the context of our
model, is a complex issue, involving assumptions not only on the
fraction of AGN bolometric luminosity emitted in this spectral
region, but also on the (distribution of) absorbing column
densities, which are expected to evolve during the SMBH build-up.
As discussed above, we  can only attempt an order of magnitude
estimate of the range of $N_H$ values from our model, but these
values are in agreement with those found by Alexander et al. Thus,
to make a crude estimate of the observable X-ray emission, we
assume as a reference  value of $N_H\sim 3 \times 10^{23}$
cm$^{-2}$, and we use for our mock objects the X-ray SED defined
by Ueda et al.\ (2003) for AGN with $\log N_H =23.5$, which
includes a Compton
reflection bumb, low energy absorbtion and exponential cut-off at 500 keV.\\

Growing AGN in SCUBA sources may contribute about 20\% to the XRB
at around 10 keV, where the contribution peaks (Fig.\
\ref{fig_xrb}). If we include only sub-millimetric sources down to
flux limit of $\simeq 1$ mJy, the contribution decreases by more
than a factor 2. These figures should be regarded as upper limits,
since we expect a fraction of AGN to be affected by column
densities larger than those found by Alexander et al. Assuming an
higher value $\log N_H =24.5$ would cause a sharp drop below 10
keV, while the estimate is little affected above this energy,
where anyway the contribution decreases.

The predicted X-ray counts due to growing SMBH in forming spheroid
are shown in Fig.\ \ref{fig_cx}, again adopting $\log N_H =23.5$.
Using $\log N_H =24.5$ instead, the predicted cumulative counts
decrease by a factor $\sim 2$ at the faint end $<10^{-15}$ erg
cm$^{-2}$ s$^{-1}$, where in any case they could provide a
sizeable fraction of sources (a few tens percent).

\section{Summary and Conclusions}
We added support to the idea that the {\it mutual} relationship
between the formation of spheroids and the AGN activity is a key
ingredient that {\it must} be included into models of galaxy
formation. The prescriptions of our ABC scenario (Granato et al.\
2001, 2004) lead to predictions in general agreement with many
observations. Here we concentrated on X-ray properties of SMGs,
which support one of its basic ingredients, namely the SF promoted
growth of a SMBH. Constraints on model details in the final phase
of this growth are expected from deeper sub-mm and X-ray
observations.

\section*{Acknowledgments}
Work supported by grant EC MRTN-CT-2004-503929. We thank an
anonymous referee for extremely useful comments.

\end{document}